\documentclass[sigconf,balance=false,authorversion]{acmart}
\AtBeginDocument{%
  \providecommand\BibTeX{{%
    \normalfont B\kern-0.5em{\scshape i\kern-0.25em b}\kern-0.8em\TeX}}}

\setcopyright{acmcopyright}
\acmYear{2023}\copyrightyear{2023}
\setcopyright{acmlicensed}
\acmConference[DataEd '23]{2nd International Workshop on Data Systems Education: Bridging education practice with education research}{June 23, 2023}{Seattle, WA, USA}
\acmBooktitle{2nd International Workshop on Data Systems Education: Bridging education practice with education research (DataEd '23), June 23, 2023, Seattle, WA, USA}
\acmPrice{15.00}
\acmDOI{10.1145/3596673.3596978}
\acmISBN{979-8-4007-0207-5/23/06}

\usepackage{balance}
\usepackage{caption}
\usepackage{subcaption}

\begin{document}

\title{Relational Playground: Teaching the Duality of Relational Algebra and SQL}
\author{Michael Mior}
\email{mmior@mail.rit.edu}
\orcid{0000-0002-4057-8726}
\affiliation{%
  \institution{Rochester Institute of Technology}
  \streetaddress{102 Lomb Memorial Drive}
  \city{Rochester}
  \state{New York}
  \country{USA}
  \postcode{14623-5608}
}

\renewcommand{\shortauthors}{Mior}

\begin{abstract}
Students in introductory data management courses are often taught how to write queries in SQL. This is a useful and practical skill, but it gives limited insight into how queries are processed by relational database engines. In contrast, relational algebra is a commonly used internal representation of queries by database engines, but can be challenging for students to grasp. We developed a tool we call Relational Playground for database students to explore the connection between relational algebra and SQL.\end{abstract}




\maketitle

\section{Introduction}

Interactive simulations have great potential to improve student learning outcomes, as has been shown with visualizations in early CS education~\cite{Guo2013}.
Students can be presented with many examples during class time but can struggle when asked to solve related problems on their own. Allowing students to explore provided problems and problems they develop independently in more detail will provide the ability to deepen understanding of the material. This ability to explore allows students to focus on their own individual understanding gaps, which may be different between students.
We aim to develop such an interactive simulation to explore relational database querying.
Although similar tools such as RelaX~\cite{Kessler2019} exist, they are not designed for novice users and do not incorporate some of the novel features we explore.

The remainder of the paper introduces our learning goals for our tool, an overview of its implementation and features, the response we have observed from students thus far, and plans for future development.

\section{Learning Goals}

While teaching an introductory database course at RIT, we observed that students faced challenges in understanding the foundations of relational database query processing.
In particular, students were taught to express queries using two methods:

\begin{enumerate}
\item Structured Query Language (SQL), which uses English keywords to express queries.
Students have little trouble understanding different components of a query in isolation, but struggle when these components are combined to form a complex query.
\item Relational algebra, which is a formal expression of database queries, is the foundation of relational database theory.
This representation is also commonly used by database query optimizers.
\end{enumerate}

When teaching both of these languages to students, it is common to show equivalent expressions in both languages to help students make connections between constructs they are familiar with in one language and those which are equivalent in the other.

When taking this approach while teaching, we found that students could understand the examples explained to them, but still struggled to formulate these equivalences on their own.
Developing a tool to allow students to explore these connections on their own provides an endless source of examples that students can tailor to address their specific gaps in understanding.
A further problem when teaching relational algebra is explaining its use in optimizing database queries.
There are multiple possible ways a query represented as SQL can be expressed in relational algebra.
Since relational algebra expressions represent specific steps which are taken when executing a query, it is possible that a database executing one expression (sequence of steps) will produce a result faster than one using another expression.
When teaching students relational algebra, especially at the graduate level, an important topic is the explanation of how database systems optimize these relational algebra expressions.

A tool which allows for exploration of the connection between SQL and relational algebra queries also presents the opportunity to show the differences between multiple relational algebra expressions which produce the same results.
By showing multiple expressions along with partial results of each step of query execution, students can explore these optimizations for themselves.
In our experience, the more examples students are presented, the easier it is for them to develop a deep understanding of the material.
Our tool enables students to explore unlimited examples since they can view relational algebra expressions corresponding to their own SQL queries.

\section{User Interface}

\begin{figure}[ht]
\centering
\includegraphics[width=0.4\textwidth]{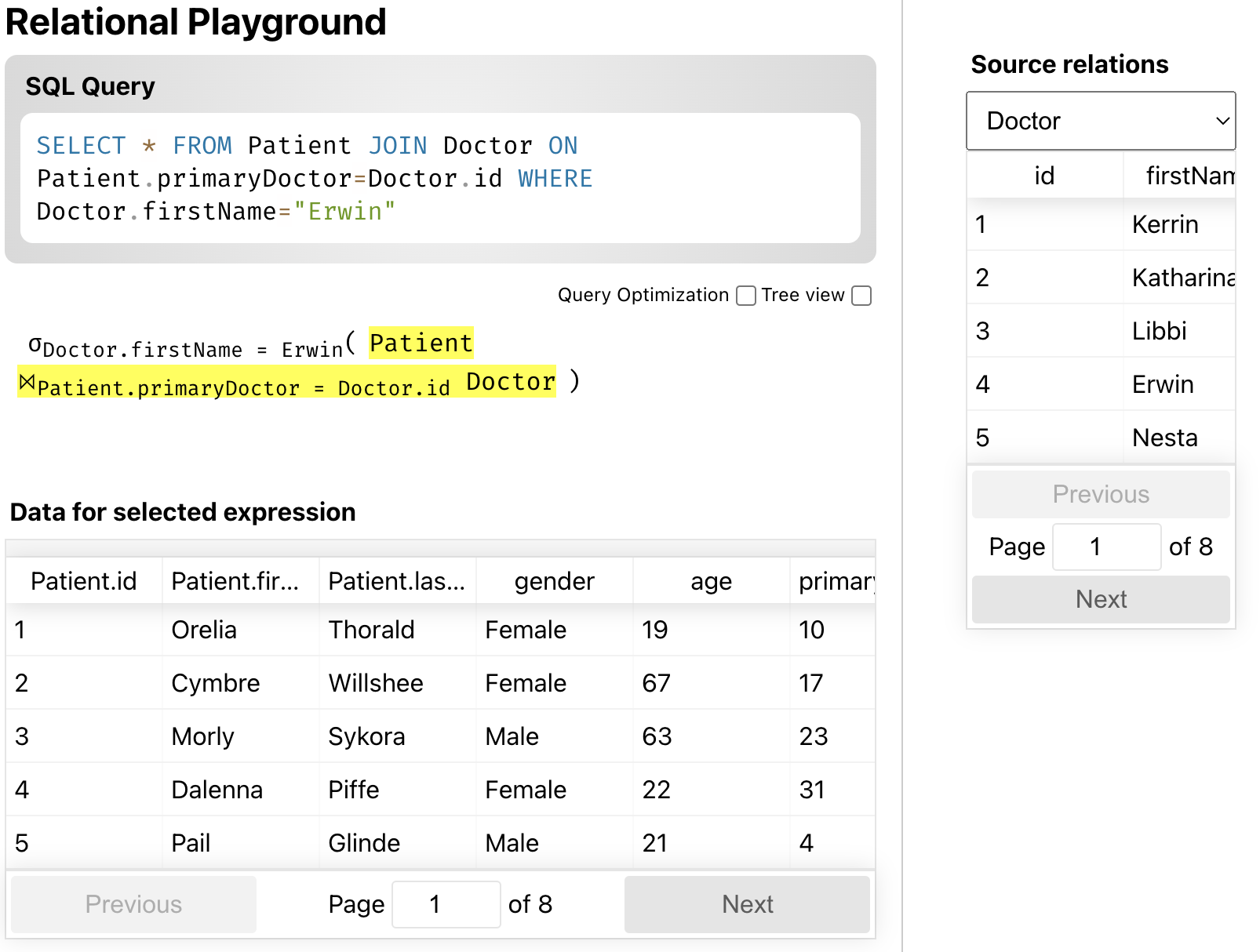}
\caption{Screenshot of the Relational Playground interface}\label{fig:interface}
\end{figure}

Our tool, which we have named Relational Playground, enables students to explore the connection between relational algebra and SQL on provided sample datasets.
The user interface for our application is shown in Figure~\ref{fig:interface}.
All queries are entered in SQL in the top input, and as the queries are modified, the corresponding relational algebra expression shown below is automatically updated.
The SQL subset supported by Relational Playground currently includes multiple joins with filtering and projection.
Our sample dataset currently contains two foreign keys, which limits the number of meaningful joins to three tables, but this can be expanded arbitrarily in the future within reasonable bounds, considering the memory and computational constraints of browser-based execution.
However, we believe that most important concepts of SQL and relational algebra can be taught with relatively small queries.

To enable students to understand how data are processed at each step in expression evaluation, they select any subexpression to highlight it and view the data at that point during query execution.
In Figure~\ref{fig:interface}, the user has the option of showing the results of executing the join before filtering has been applied.
If a student is unclear of the effect of a relational algebra operator, viewing the data before and after its application may be helpful in clarifying understanding.
For example, in Figure~\ref{fig:interface}, a student may choose to click the selection operator, which will update the data displayed below to show only selected tuples from the join.
The source relations available for use in queries can be browsed on the right of the interface.

This browser-based tool is currently available online\footnote{\url{https://www.relationalplayground.com}} and the source is in active development and available on GitHub\footnote{\url{https://github.com/dataunitylab/relational-playground}}.

\section{Query Optimization}

\begin{figure}
\centering
\includegraphics[width=0.4\textwidth]{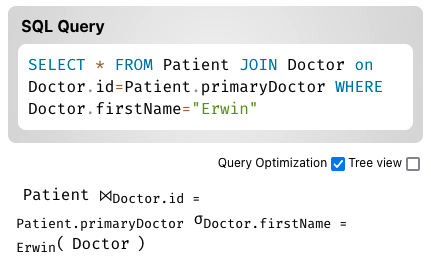}
\caption{Post-optimization view in Relational Playground}\label{fig:query_opt}
\end{figure}

One of the unique features of Relational Playground is the ability to support logical query optimization.
Since the entire environment runs on a small dataset in the user's Web browser and not on a server, we do not attempt to optimize based on query runtime, since it can be unpredictable in this environment.
We also attempt to simplify the optimization process by avoiding optimizations that rely heavily on statistics.
Instead, we focus on common heuristics that can demonstrate the intuition behind optimization.
In doing so, we also avoid considering the time required to run the optimizer itself.
Since our focus is on fairly simple queries, this has not proven to be an obstacle.
The first optimization we focus on is predicate pushdown past joins, as our experience has shown that this is the simplest possible explanation for students to understand.
This pushdown tends to result in a more optimal query plan, independent of any statistics on the tables involved.

\begin{figure}[ht]
\centering
\includegraphics[width=0.3\textwidth]{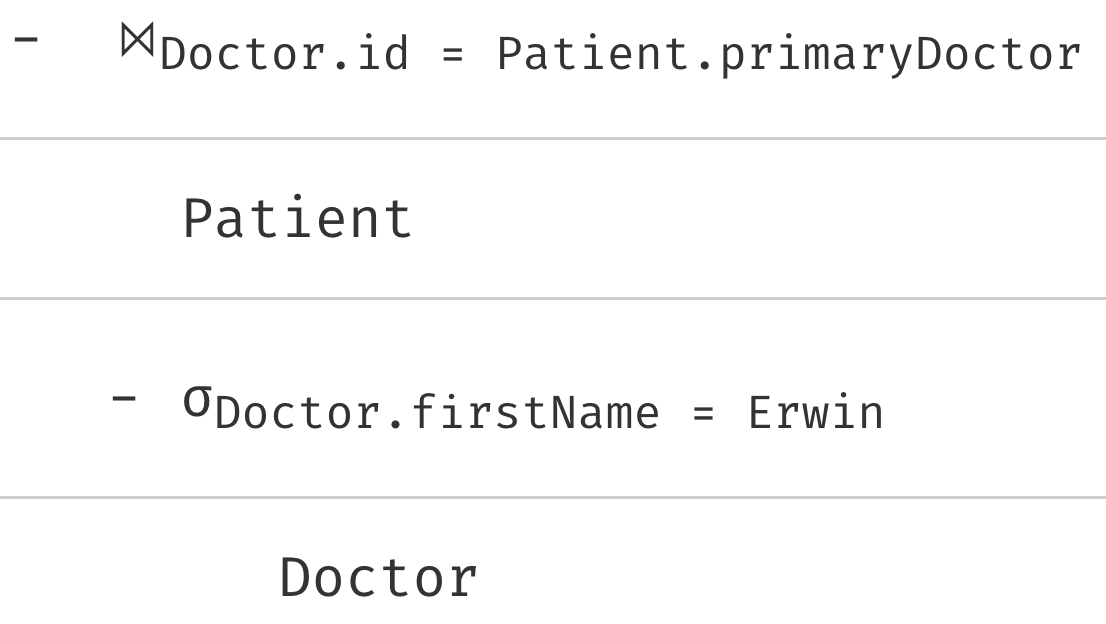}
\caption{Tree view of optimized relational algebra expression}\label{fig:query_tree}.
\end{figure}

Our rule for predicate pushdown simply matches a filter which appears after a join and checks for a predicate involving only one of the relations as input to the join.
If such a match is found, the predicate is moved past the join.
Although this is a simple optimization, the ability to enter any query and see possible effects allows students to develop intuition about how the optimization works.
This is similar to the database optimization curriculum presented by Davis~\cite{Davis2022} but we are able to automate the optimization step.
Although this eliminates the need for students to perform the optimization manually, we believe that the added exploratory element is beneficial to student learning.
To explore the optimization process, students can toggle optimization as shown in Figure~\ref{fig:query_opt}.
Since optimization of relational algebra is often taught in the form of tree transformations, we also allow students to view the expression formatted as a tree as in Figure~\ref{fig:query_tree}.
Toggling query optimization on and off in tree view allows students to easily see how the expression tree is transformed by optimization.

\section{Student Response}

We piloted this tool in a section of Introduction to Big Data, which is a required course for computer science graduate students who wish to take any further classes in data science.
Of the students who interacted with the tool and completed the survey ($n=10$), 80\% claimed that the tool was at least somewhat useful in improving their understanding of relational algebra.
Most of these students used the tool to study for tests, and many used it multiple times throughout the semester.
We note that this survey was performed before the implementation of query optimization, as described in the previous section.
We plan to send a revised survey to students who interact with the updated version of the tool and seek more detailed feedback.

\section{Related Work}

As mentioned above, the RelaX~\cite{Kessler2019} relational algebra calculator is the closest to our tool.
RelaX also allows students to enter SQL queries and view the corresponding extended relational algebra expression.
The subset of SQL currently supported by RelaX is larger than supported by Relational Playground due to its use of extended relational algebra, but a key difference is that RelaX does not support query optimization.
We believe that our query optimization support is particularly useful for understanding how relational algebra is used in practice by query optimizers.
The Umbra demonstration interface~\cite{Neumann20} allows users to view the steps in query optimization, but focuses on describing the specific capabilities of its particular engine rather than providing concise examples.

Taking the opposite approach of Relational Playground, RA~\cite{radb} allows the conversion of relational algebra expressions to SQL queries.
RA has been used for tasks such as presenting small counterexamples for incorrect SQL queries~\cite{Miao19}.
In our setting, converting relational algebra to SQL could prove useful for enabling bidirectional editing of relational algebra and SQL.

\balance

\section{Future Work}

Relational Playground currently supports a limited subset of SQL, including nested select-project-join (SPJ) queries corresponding to basic relational algebra operators.
We would like to expand the supported SQL subset to include queries, aggregation, grouping, sorting, and basic arithmetic.
Other possible extensions include temporal relational algebra~\cite{Grandi17} to support streaming SQL.
We are interested in exploring the creation of an editor for relational algebra expressions to enable bidirectional conversion between SQL and relational algebra.
This will allow students to better understand the connection between these two languages.

We plan to add more heuristic query optimizations to Relational Playground.
For example, Apache Calcite~\cite{Begoli2018} contains more than 100 optimization rules.
Although many of these are likely too complex for our setting, we expect that several of these rules will prove useful.
Furthermore, each rule is currently applied without explanation.
Users can see expressions before and after optimization, but there is no indication of how optimizations are applied.
It would be helpful to provide the opportunity to toggle individual optimization rules, as well as show how each rule matches the original relational algebra expression and how the optimized expression is generated.

In the future, it is also expected that this tool could be expanded to make connections to other non-relational database systems.
For example, when teaching the connection between NoSQL and NewSQL systems, which are evolutions of the relational databases students have already studied, many operators similar to those used in relational algebra are applicable in these contexts.
This aspect of future work is similar to TriQL~\cite{Alawini22} which allows users to view queries for databases using a variety of data models, but does not include relational algebra or query optimization.
Future extensions to this tool could be extremely useful in this context.

Finally, we would like to better evaluate the usefulness of Relational Playground in improving student understanding.
We plan to include both additional surveys with a larger sample size and a controlled study in which students are assessed on relevant learning outcomes with and without access to our tool.
In addition, we plan to interview students to determine possible usability issues and improvements.

\begin{acks}
This work was funded by the Provost’s Learning Innovations Grants (PLIG) program at the Rochester Institute of Technology.
Special thanks to Carson Bloomingdale and Aryan Jha, who spent significant time on development.
\end{acks}

\bibliographystyle{ACM-Reference-Format}
\bibliography{references}


\end{document}